\documentclass[preprint]{revtex4}

\usepackage{amsmath}	
\usepackage{amssymb}
\usepackage{graphicx}	
\usepackage{epstopdf}	
\usepackage[utf8]{inputenc}
\usepackage{subfigure}
\usepackage{hyperref} 
\usepackage{placeins}

\newcommand{\figref}[1]{{Fig.~\ref{#1}}}
\newcommand{\subfigref}[2]{{Fig.~\ref{#1}~#2}}

\newcommand{\forref}[1]{{(\ref{#1})}}
\newcommand{\longforref}[1]{{Eq.~(\ref{#1})}}

\newcommand{\eqns}[0]{{Eqns.\ }}

\newcommand{\var}{\textrm{Var}}

\newcommand{\MQ}{\overline{M}}
\newcommand{\SQ}{\overline{S}}
\newcommand{\pQ}{\overline{p}}

\newcommand{\mysection}[1]{\section*{\hspace{-\parindent}#1}}

\setcitestyle{round}
\begin{document}


\title{Unstable Price Dynamics as a Result of Information Absorption in Speculative Markets}

\author{Felix Patzelt}
\email{felix@neuro.uni-bremen.de}
\author{Klaus Pawelzik}
\affiliation{Institute for Theoretical Physics, University of Bremen, Germany}
\date{\today}

\maketitle


\mysection{Abstract}

\noindent\textbf{%
In speculative markets, risk-free profit opportunities are eliminated by traders exploiting them.
Markets are therefore often described as ``informationally efficient'', rapidly removing predictable price changes, and leaving only residual unpredictable fluctuations. This classical view of markets absorbing information and otherwise operating close to an equilibrium is challenged by extreme price fluctuations, in particular since they occur far more frequently than can be accounted for by external news.
Here we show that speculative markets which absorb mainly self-generated information can exhibit both: evolution towards efficient equilibrium states as well as their subsequent destabilization.
This peculiar dynamics, a generic instability arising from an adaptive control which annihilates predictable information, is realized in a minimal agent-based market model where the impacts of agents' strategies adapt according to their trading success. This adaptation implements a learning rule for the market as a whole minimizing predictable price changes. The model reproduces stylized statistical properties of price changes in quantitative detail, including heavy tailed log return distributions and volatility clusters. 
Our results demonstrate that the perpetual occurrence of market instabilities can be a direct consequence of the very mechanisms that lead to market efficiency.
}\\


\noindent Social systems self-organize. In consequence, collective behaviors can emerge that appear to pursue a common goal which is actually not present in the behavior of the single agents \cite{smith1776wealth}. The view that markets in fact operate in distinguished equilibrium states became influential in economics \cite{samuelsonEconomics}. Here, a fundamental hypothesis is that markets operate  informationally optimal. That is, prices are assumed to "fully reflect available information" \cite{fama1970efficientMarkets}, or at least come close to this limit \cite{grossmann1980efficient}, such that risk-free (arbitrage) profits cannot be made by \mbox{(re-)using} said information. If true, one of the implications of this "Efficient Market Hypothesis" (EMH) is that resulting prices fluctuate randomly  \cite{samuelson_prices}.

\begin{figure}
	\includegraphics[width=\textwidth]{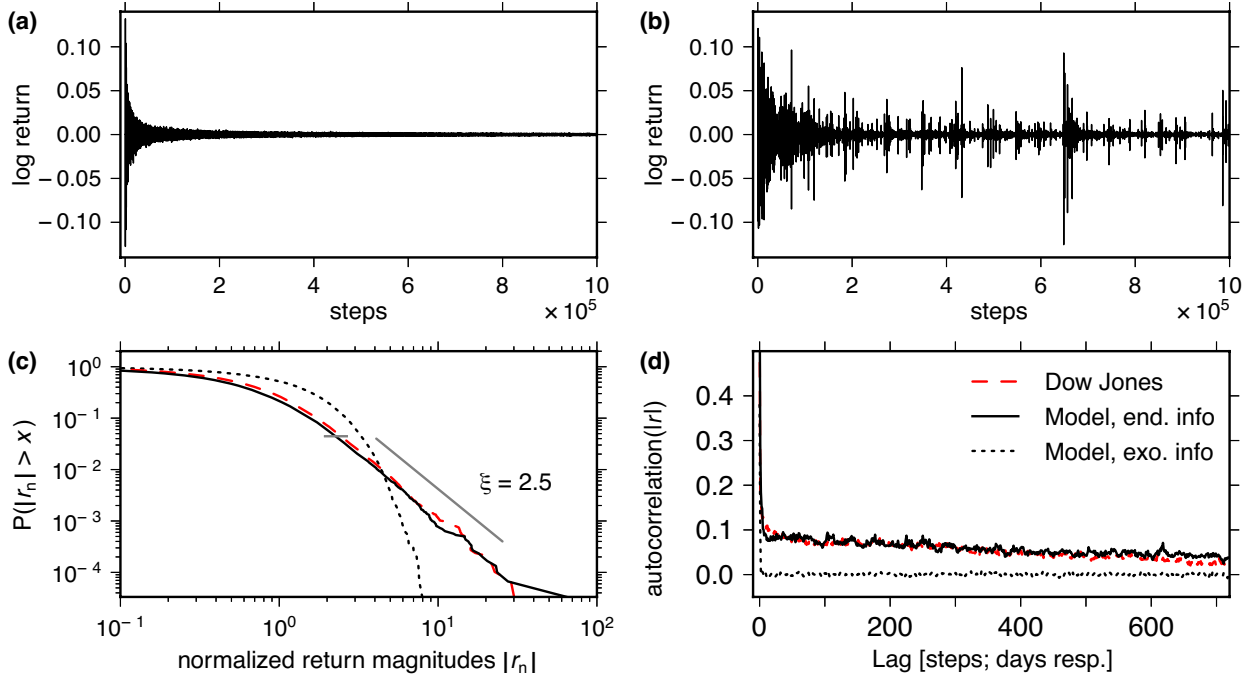}
	\caption{Analysis of log returns $r(t+1)=\log{p(t+1)}-\log{p(t)}$, where $p(t)$ is the price at time step $t$. (a): Time series of the trading model with uniformly distributed exogenous information. Parameters: $N_s = 2^{10}$, $N_p = 0$, $D = 2^9$, $\gamma = 0.8$. (b): Time series for the same model, but with endogenous information. (c): Complementary cumulative distribution function of log return magnitudes obtained by rank ordering. Dotted black: The same simulation as in (a). Solid black: the same simulation as in (b). Dashed red: Daily returns for the Dow Jones Industrial Average (DJIA) since 1900. Short grey line: Hill estimator for the scaling exponent $\xi$ of the black curve, where large returns show power-law scaling $P(|r| > x) \propto x^{-\xi}$. The cutoff optimizes the Kolmogorov–Smirnov statistic \cite{patzelt2011critical}. For a fair comparison, $3\cdot10^4$ time steps after transients were analyzed for the simulations to match the length of the Dow Jones time series. Each time series was normalized by dividing by its standard deviation. (d): Autocorrelations of the log-return magnitudes $|r|$. Line styles are identical to (c). Positive autocorrelations for the model with endogenous information, and the DIJA persist for more than 2 years.}
	\label{fig:transient}
\end{figure}

Empirical findings in favor of the Efficient Market Hypothesis include the general absence of exploitable autocorrelations among price changes in financial markets \cite{Fama1998EfficiencyReturnsBehaviour}. The magnitudes of price changes ("volatilities"), however, are correlated for long periods of time. That is, large price changes are typically followed by large ones and small changes by small ones, indicating that markets are complex dynamical systems involving long memories \cite{Mandelbrot1963Prices,  Boucheaud2009MarketsSlowlyDigest}. Furthermore, logarithmic price changes (log returns) exhibit large fluctuations even in the absence of external news \cite{cutler1989prices, joulin2008news}, contradicting naive expectations for systems in stable equilibria. More precisely, log return distributions exhibit heavy tails where events that are many times bigger than the standard deviation occur at a much higher frequency than what would be expected if they were Gaussian distributed \cite{Mandelbrot1963Prices, farmer99scaling} (\subfigref{fig:transient}{(c)}). These so-called stylized facts of price time series hint at underlying dynamics that are substantially self-referential \cite{bouchaud2008loops}. They were associated either with herding effects \cite{devenow1996herding}, ``bubbles", or with the interactions of heterogeneous traders with limited rationality \cite{lux1999multi-agent} in a market exactly at a critical point \cite{challet2004market_chapter}.

Here we investigate whether the apparent antinomy of stabilizing information efficient control and dynamics resembling systems operating close to criticality can be resolved by a recent non-economic theory \cite{patzelt2011critical}: It was shown that adaptive control of a dynamical system can generically lead to an instability where the susceptibility to noise dramatically increases close to the point of perfect balance. This principle applies to markets if two requirements are fulfilled: First, markets have to absorb information about predictable price changes. Ideally, this property should hold independently of the rationality of the individual traders, which cannot be guaranteed. Second, a self-referential market needs to become susceptible to residual noise once all locally relevant information has been exploited.

The first property is a rather common view in economics. If a profit opportunity is present, investors will increasingly exploit it until their price impacts cancel said opportunity. For example, if a certain stock is priced too low, traders will increase buy orders and thereby raise its price. The second property is also  intuitive. As traders try to detect trends or patterns in the price dynamics, they effectively predict how the market reacts to available information. However, once the agents' actions have led to a balanced equilibrium, it becomes increasingly difficult to distinguish predictable price fluctuations from random noise. If traders then act upon the random fluctuations as if they would hold meaningful new information, their actions will not be balanced anymore. That is, it may be impossible to predict whether a group of traders overreact to the supposed new information and to attenuate the resulting price jump by exploiting it. Therefore, atypically large price movements may become much more likely than expected for a Gaussian distribution.

\mysection{The Model}

As a concrete example of the fundamental dynamical instability arising from information absorption as it may be realized in financial markets, we introduce a minimal agent-based trading model. Each agent is representative of one trading strategy and possesses two types of assets which are called money and stocks in the following. For simplicity, we consider a coarse-grained price time series where one step could be considered as e.g. a day. In each time step, every agent either buys or sells an amount of stocks. The fraction of an agent's money that is used to buy stocks or vice versa is denoted by the use parameter $\gamma$. At each time $t$, all trades are performed at the same price $p(t)$ which is determined from the ratio of total demand and supply. Hence, trading per se conserves the total amounts of each asset, money and stocks.

Agents base their decisions on public information states. In each time step $t$ one of $D$ possible states, which is denoted by an index $\mu(t) \in \{1, \dots, D\}$, is conveyed to the agents. Each agent's   decision to either buy or sell is fixed in time for each $\mu$ and assigned at random. We consider two different methods for the generation of these information states:

For endogenous information, agents possess a memory of the most recent $K$ signs of the log returns which indicate whether the prices $p(t-K), \dots,  p(t)$ decreased or increased with respect to their predecessor. This information can take one out of $D = 2^K$ possible states.

For exogenous information, $\mu(t)$ are drawn randomly and independently with probability $P_\textrm{ext}(\mu)$. Unless stated otherwise all $\mu$ have equal probabilities $P_\textrm{ext}(\mu) = 1 / D$. We also investigated mixed information and obtained results similar to the endogenous case (see supplement).

We focus on markets that are dominated by speculators who can win or loose assets only by betting on price changes within the market. To investigate the effect of a small number of traders that convey new assets to the market or draw out their profits, we divide the agents into $N_s$ speculators and $N_p$ producers. Producers' resources by definition stay constant. Thus, speculators only redistribute their wealth and are competitive whereas producers provide a predictable supply of liquidity and stocks. All agents are initially provided with equal amounts of assets.

\mysection{Stylized Facts}

Log-return time series for the model with exogenous information are shown in \subfigref{fig:transient}{(a)}. A strong reduction of initial fluctuations is observed leaving only a narrow band of Gaussian distributed returns after the transient. \subfigref{fig:transient}{(b)} shows the endogenous case. Here, in contrast, initial return magnitudes are reduced only in the mean. The magnitudes of the few most extreme returns, however, are less reduced. The remaining fluctuations are analyzed in \subfigref{fig:transient}{(c)}, where cumulative distributions of return magnitudes are shown for both cases and compared to the Dow Jones Industrial Average (DJIA). The latter serves as an example for a typical price time series. For the endogenous case, the distribution tail is well described by a power-law and in good agreement with the DJIA. Return fluctuations in the endogenous case also tend to form clusters in time. This effect is quantified by long-range temporal correlations of return magnitudes shown in \subfigref{fig:transient}{(d)} and is also consistent with the DJIA.

\begin{figure}[h!]
	\includegraphics{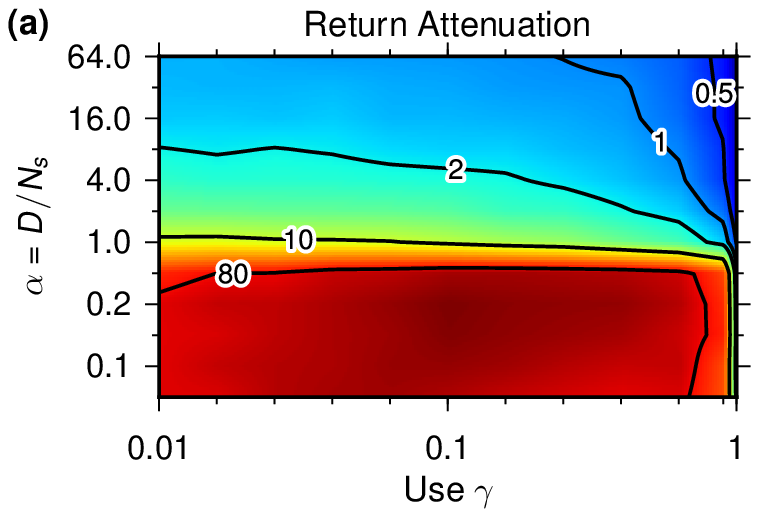}\\[.5\baselineskip]
	\includegraphics{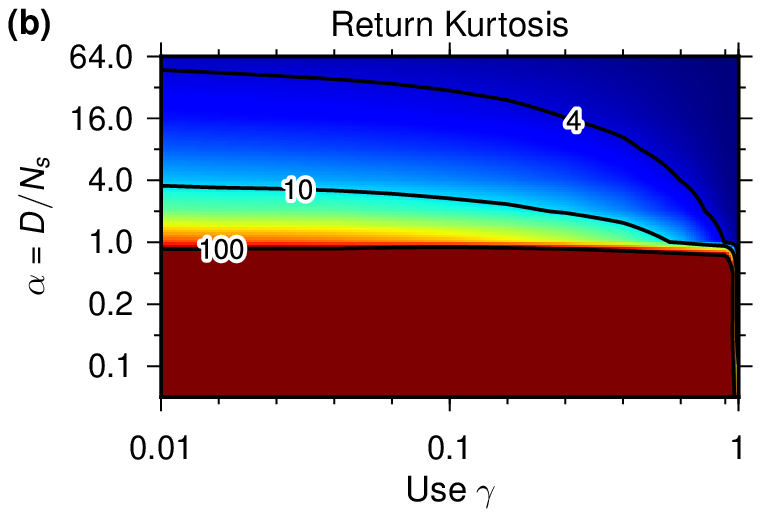}
	\caption{\textbf{(a)}: reduction of average return magnitudes during transients and \textbf{(b):} kurtosis of log-returns of the model with endogenous information for different $\alpha$ vs $\gamma$. A kurtosis of $3$ corresponds to a normal distribution. Here, the system size is set constant at $N_s = 2^{10}$ speculators and $N_p = 2^4$ producers. For each time series, $2 \cdot 10^7$ time steps were simulated. Reductions are measured as the ratio between the mean log-return magnitudes during the first $10$ and the last $10^7$ timesteps. The kurtoses were calculated for the last $10^7$ timesteps. Simulations were performed on a grid, all axis ticks correspond to node positions. For each node on the grid, $50$ time series were analyzed and results were averaged. Values in between nodes were obtained by linear interpolation after logarithmizing the values at each node. Color mapping was performed on the log-values, contour line labels are the actual values for variances and kurtoses. For $\alpha \leq 1/2$, the kurtoses reach extreme values that can not be reliably estimated from finite time series. Therefore, the color scale in (b) was set to not extend to values above 100.}
	\label{fig:use_vs_bet}
\end{figure}

\mysection{Dynamics with Exogenous Information}
To understand the model dynamics, we first consider the exogenous case which is fully analytically tractable. The rules of asset redistribution by trading are equivalent to a learning rule related to gradient descent, allowing for the market as a whole to minimize predictable price changes. The reason for this stabilizing control is that trading success increases the impact of agents whose actions contribute to a reduction of price fluctuations. A phase transition with respect to the critical parameter $\alpha = D / N_s$ is identified at $\alpha = 1/2$, the point where random binary vectors (the agents) with positive weights (the assets) form a complete basis in the $D$-dimensional strategy space in the limit $N_s \rightarrow \infty$. Beyond this point, a speculative market without producers evolves the distribution of assets onto a manifold where the price is invariant to trading. That is, agents still trade and exchange assets, but the price remains constant. Markets that include  producers still exhibit finite returns also for $\alpha < 1/2$. Otherwise, for $N_p \ll N_s$, return distributions only depend weakly on $N_p$. See \figref{fig:use_vs_bet}, and the supplementary material for more details.

\mysection{Information Annihilation Instability}

\begin{figure}
	\includegraphics{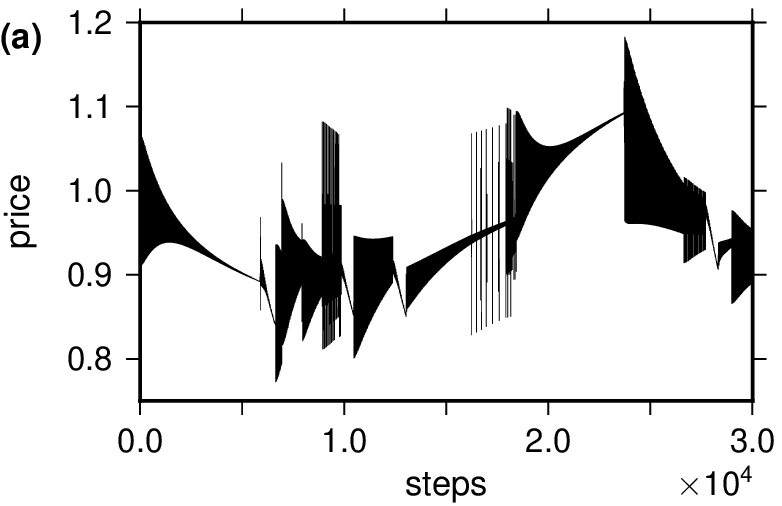}\\[.5\baselineskip]
	\includegraphics{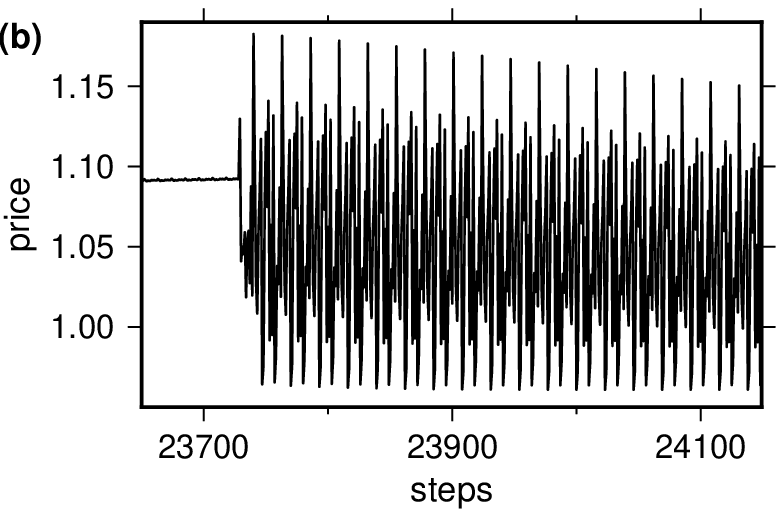}
	\caption{\textbf{(a)}: Price time series with a very slow rate of resource redistribution (use) $\gamma = 0.01$. Other parameters: $N = 2^{10}$, $N_p = 2^4$, $D = 2^9$. \textbf{(b)}: A zoom in on the time series shown in (a). }
	\label{fig:price_slow}
\end{figure}

When the $\mu(t)$ are endogenously generated, the same mechanism of information absorption present in the exogenous case ensures that the system relaxes towards local price equilibria and returns vanish, but only transiently. \subfigref{fig:price_slow}{(a)} shows the price time series for a simulation with a very small use $\gamma$ which formally is a learning rate. At any point in time, the system moves towards a certain price which characterizes a local equilibrium. As the system approaches this equilibrium, price fluctuations are reduced. These fluctuations generally consist of complex oscillations like the one shown in \subfigref{fig:price_slow}{(b)}. The equilibria become unstable once all predictable information is exploited by the speculators. Then, even little overshooting of the adaptation dynamics or noise can lead to price changes corresponding to information states that were not predicted by patterns in the immediate past. Because the market is not well-adapted to these new states, the possibility of large price changes increases dramatically. Compared to \cite{patzelt2011critical}, we here discovered an instability due to information annihilation in a mathematically different way, which demonstrates that this concept is even more general.


For larger $\gamma$, this behavior is not as obvious: time series appear random and distinct oscillations are rarely visually recognizable (\subfigref{fig:transient}{(b)}). Still, phase diagrams from extensive simulations demonstrate that return distributions are largely unaffected by these effects over wide ranges of $\gamma$: \subfigref{fig:use_vs_bet}{(a)} shows how much initial mean log return magnitudes are reduced for different $\alpha$ and $\gamma$; \subfigref{fig:use_vs_bet}{(b)} shows the impacts that infrequent extreme returns have on the remaining variances, which are measured by kurtoses after transients.
The stronger the reduction of return magnitudes (\subfigref{fig:use_vs_bet}{(a)}), the heavier tailed the return distributions are (\subfigref{fig:use_vs_bet}{(b)}). Therefore, a clear link between information annihilation and extreme returns is established in our model for the whole parameter space. This relation between return reduction and kurtosis is not found for exogenous information drawn from a uniform distribution (\subfigref{fig:transient}{(b)}, and supplement).

\FloatBarrier

\mysection{Surprising Information Causes Large Returns}

\begin{figure}
	\includegraphics{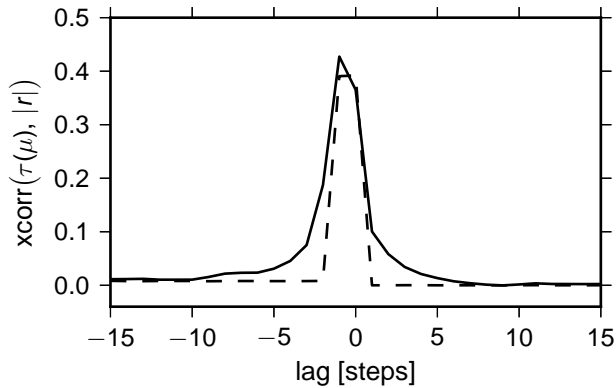}
	\caption{Correlation of return magnitudes $|r|$ with the time $\tau$ since the corresponding information states occurred last. Model parameters: $D = 2^{10}$, $N_s = 2^{11}$, $N_p = 0$, $\gamma = 0.5$. Simulation length:  $T=2\cdot10^7$; the first $T/2$ timesteps were discarded for the analysis. Black line: Endogenous information. Dashed line: Exogenous information with $P_\textrm{exo}(\mu) \propto \exp(-0.02\, \mu)$, leading to $P(\tau) \propto \tau^{-2}$. Both lines are averages over $10$ simulations.}
	\label{fig:surprise}
\end{figure}

We find that large returns are caused by information states that have not occurred for a long time. Intuitively, such states are more surprising and therefore carry more information than the ones visited more  recently -- a concept that is closely related to Shannon information. The more surprising an information state is, the higher the corresponding log return. The correlation between log return magnitudes and the times $\tau$ that have passed since the respective information states occurred last is shown in \figref{fig:surprise} (solid line) for endogenous information. Here, the absorption of local information in combination with rare jumps leads to a strongly inhomogeneous distribution of visiting frequencies over the information set: the probability distribution $P(\tau)$ is power-law tailed with an exponent of approximately $2.5$ (see supplement). 
This suggests that the self-reflexive dynamics for endogenous information generates a characteristic distribution of information states that ultimately underlies extreme price fluctuations. We tested this hypothesis by using inhomogeneously distributed exogenous information states leading to similarly distributed $\tau$. Then, as in the endogenous case, return magnitudes are strongly correlated with $\tau$ (\figref{fig:surprise}, dashed line).

\mysection{Discussion}
Our model demonstrates that even a simple yet plausible mechanism of order size adaptation suffices to coordinate the impacts of diverse strategies such that information becomes completely absorbed in the price. The proof that the trading rules in our model correspond to an efficient gradient based learning rule that minimizes predictable return magnitudes provides a rigorous link of a fundamental market mechanism to adaptive control. We tested variations of this mechanism including adaptations of individual agent use parameters, and order-book based pricing \cite{Rothenstein_2003a}; information absorption proved to be a robust property. We therefore consider it highly plausible to prevail in real markets despite the complexities of real pricing mechanisms and order size adaptations, as long as the latter correlate with trading success.

``Rationality'', in the sense of an efficient adjustment of prices to new information, in our model emerges from the collective behavior of many traders. We find that a minority rule with respect to the returns is a dominating factor in its dynamics. That is, traders whose actions counteract those of the majority profit most.

In the well known minority game \cite{challet2004market_chapter} (MG), however, adaptation is based on the choices of the participating agents, and a single-step payoff with respect to the excess demand. This differs from our model where the market adapts traders' impacts based on price changes, and also over different time scales.
Most importantly, the mechanism for herding previously discussed for MGs is a breakdown of the efficient coordination of agents in overcomplete markets where they become too correlated. This leads to an increase of average fluctuations and does not depend on how the information conveyed to the agents is generated. 

In contradistinction, the instability due to adaptive control is independent of microscopic interactions and in fact was first realized in a one-dimensional random map \cite{patzelt07}. The heavy tailed distributions here are a direct consequence of the elimination of local trends or patterns which yields a net decrease in average fluctuation magnitudes. These fluctuations are therefore a sign of high efficiency and do not signal its breakdown: they reflect surprising information. Furthermore, a closed loop involving endogenous information is essential for creating this instability leading to extreme fluctuations that are not caused by external events. Therefore, the results presented in this paper are complementary to the published findings based on MGs in several respects.
It is, however, possible to formulate a minority game corresponding to our model, but that would go beyond the scope of this paper.

The fact that efficient information annihilation does not result in a unique and stable equilibrium, but instead can lead to local states that perpetually become unstable, provides a new and comprehensive solution to a long standing antinomy in economics where both notions have been considered mutually exclusive \cite{lux1999multi-agent}. Furthermore, our adaptive agent-based model not only demonstrates the theoretical principle of information annihilation instability, but can reproduce important ``stylized facts'' of financial markets also quantitatively to a surprising extent. The parameter sets that best reproduce the data correspond to situations where agents on average profit most (see supplement). This is plausible because real traders can choose among different markets. These findings suggest that the information annihilation instability is indeed one of the main causes of the notorious large jumps in price time series, particularly in strongly speculative markets. An empirical confirmation of our theory would require identifying the information states that cause large price changes in a given market, for which it then makes directly testable predictions.



\bibliography{references}

\vspace{\baselineskip}

\vspace{\baselineskip}
\noindent\textbf{Acknowledgments.} We thank Josephine Mielke who developed a precursor model to the one presented in this paper, Andreas Kreiter for his help improving this letter's general understandability, and the Volkswagen Foundation for funding Felix Patzelt. Dow-Jones Data from: Department of Statistics at Carnegie Mellon University; download: http://www.analyzeindices.com/dowhistory/djia-100.txt.

\vspace{\baselineskip}
\noindent\textbf{Author Contributions.} The authors contributed equally to this work.


\newpage

\def\thefigure{S\arabic{figure}}
\setcounter{page}{1}
\pagenumbering{roman}

{
\begin{center}
	{\Large Supplementary Materials for}
	
	\vspace{.5\baselineskip}
	{\large Critical price dynamics as a result of information annihilation\\ in speculative markets}
	
	\vspace{.5\baselineskip}
	\noindent Felix Patzelt and Klaus Pawelzik
	
	\noindent correspondence to felix@neuro.uni-bremen.de
\end{center}

\vspace{\baselineskip}
\mysection{Contents}

\renewcommand{\labelitemi}{}

\begin{itemize}
\item Supplementary Methods
	\begin{itemize}
		\item Model
		\item Invariant Manifold
		\item Completeness of the Strategies
		\item Gradient Descent
	\end{itemize}
\item Supplementary Figures
	\begin{itemize}
		\item Phase Diagrams for Exogenous Information
		\item Speculators and Producers
		\item Distribution of Information Ages (Surprise)
		\item Mixed Information
	\end{itemize}
\item Supplementary Discussion: Income and the Critical Point
\end{itemize}
}

\newpage

\section{Supplementary Methods}

\subsection{Model}

\noindent Each agent $i = 1, \dots, N$ possesses two types of assets which without loss of generality are called money $M_i(t)$ and stocks $S_i(t)$. N is the total number of agents. In each round, every agent places a market order to either buy or to sell an amount of stocks. Since trading should conserve the amount of traded assets, the price in each round is determined by:
\begin{equation}
	p(t) = \frac{\delta(t)}{\varsigma(t)}
	\label{eq:price}
\end{equation}
with demand $\delta$ and supply of stocks $\varsigma$. This is a fair rule that could be used in a real market with only market orders.

For a market including stochastic limit orders gathered over some period of time, consider the hypothetical price $p^*(t)$ at which trades would take place if all agents scaled their orders by a common factor. Then the volume would change, but to preserve market clearing the price is not affected; that is $p(t)^* = p(t)$. Therefore, the price is a function of the ratio of demand and supply. After linearization of this function for small small excess demands the price is proportional to the aforementioned ratio which justifies this choice of the pricing rule also as an approximation of the mean prices obtained from limit orders.

Agents base their decisions on a public information state. In each time step one of $D$ possible states, which is denoted by an index $\mu(t)$, is conveyed to the agents. We distinguish two different methods for the generation of these information states at each time step $t$:

For exogenous information, $\mu(t)$ are independent identically distributed random variates drawn from a distribution $P_\textrm{ext}(\mu)$.

For endogenous information, agents possess a memory of the most recent $K$ signs of the log returns. To eliminate the possibility of a lock, signs for vanishing returns are chosen at random. This information can take one out of $D = 2^K$ possible states:
\begin{equation}
	\mu(t+1) = \big(2\, \mu(t) + \Theta(r(t) + \eta) \big)\! \mod D
	\label{for:mu}
\end{equation}
where $\Theta$ denotes the Heaviside step function and $\eta$ is an arbitrarily small symmetric random variable with zero mean. Simulation results do not depend on $\var(\eta)$ as long as it is small enough.

Each agent's $i$ consequent decision is now determined by a strategy vector whose elements $\sigma_i^{\mu}$ are drawn randomly out of $\left\lbrace0,1\right\rbrace$ once and then kept constant. These two possible decisions correspond to trading an amount $m_i(t)$ of money or an amount $s_i(t)$ of stocks for the respective other asset in the next round. Orders are placed with a constant use parameter $\gamma$:\\

Case $\sigma_i^{\mu(t)} = 1$ (agent $i$ buys stocks):
\begin{eqnarray}
	m_i(t) &=& \gamma\, M_i(t)\\
	s_i(t) &=& 0\nonumber
\end{eqnarray}

Case $\sigma_i^{\mu(t)} = 0$ (agent $i$ sells stocks):
\begin{eqnarray}
	m_i(t) &=& 0\nonumber\\ 
	s_i(t) &=& \gamma\, S_i(t)
	\label{for:si}
\end{eqnarray}
Demand and supply are the sums of all buy and sell orders respectively
\begin{eqnarray}
	\delta(t) &=&\sum_{i=1}^{N} m_i(t) + \epsilon\\
	\varsigma(t) &=& \sum_{i=1}^{N} s_i(t) + \epsilon
	\label{dem_sup}
\end{eqnarray}
where $\epsilon$ is a small positive number. This ensures that prices and returns are always well defined. The cases with zero demand or supply are, however, irrelevant for all practical purposes. A sufficiently small $\epsilon \ll 10^{-3}$ does not influence simulation results to a meaningful degree. All figures were generated using $\epsilon = 10^{-10}$. 

We investigate the effect of a market ecology by dividing the agents into $N_s$ speculators and $N_p = N - N_s$ producers. We focus on markets that are dominated by speculators whose resources are redistributed due to trading:
\begin{equation}
	\left. \begin{array}{lcl}
		M_k(t+1) &=& M_k(t) - m_k(t) + s_k(t)\, p(t)\\
		S_k(t+1) &=& S_k(t) - s_k(t) + m_k(t) / p(t)
	\end{array}\right\} N_p < k <= N.
	\label{eq:redistribute}
\end{equation}
Producers' resources stay constant throughout the game:
\begin{equation}
	\left. \begin{array}{lcl}
		M_j(t) &=& M_j(0)\\
		S_j(t) &=& S_j(0)
	\end{array}\right \}\quad j < N_p.
	\label{eq:redistribute_p}
\end{equation}
Thus, only speculators redistribute their wealth and are competitive whereas producers provide a predictable supply of liquidity and stocks. All agents are initially provided with equal amounts of resources $M_i(0)=S_i(0)= 1$.

\subsection{Invariant Manifold}

We show, that if one distribution of resources $(\MQ, \SQ) = (\MQ_1, \dots, \MQ_N, \SQ_1, \dots, \SQ_N)$ exists for which the price $p(\MQ, \SQ, \mu) = \pQ$ is independent of the information $\mu$, this price is invariant with respect to any resource redistribution due to trading in a purely speculative market. That is, there is a manifold $Q = \{(\MQ', \SQ')\ |\ p(\MQ', \SQ', \mu) = \pQ\ \forall\ \mu\}$ of distributions of stocks for which the price is independent of $\mu$ and this manifold is closed with respect to trading according to \longforref{eq:redistribute}. For the proof, assume that at some point in time the system is in a suitable state such that
\begin{eqnarray}
	\frac{\delta(\MQ, \mu)}{\varsigma(\SQ, \mu)} &=& \pQ \quad \forall\ \mu(t) \Leftrightarrow\\[.5\baselineskip]
	\delta(\MQ, \mu) - \pQ\, \varsigma(\SQ, \mu) &=& \nonumber\\
	\gamma \sum_{i=1}^{N} \left( \sigma_i^{\mu}\, \MQ_i - \pQ\,  (1-\sigma_i^{\mu})\, \SQ_i \right) &=& 0  \quad \forall\ \mu(t).
	\label{for:steady_state_condition}
\end{eqnarray}
Then, denoting the distributions of stocks and money after trading by $\MQ_i'$ and $\SQ_i'$ we obtain:
\begin{align}
	&\frac{1}{\gamma} \left(\delta(\MQ', \mu') - \pQ\, \varsigma(\SQ', \mu')\right) = \sum_{i=1}^{N} \sigma_i^{\mu'}\, \MQ_i' - \pQ \sum_{i=1}^{N} (1-\sigma_i^{\mu'})\, \SQ_i'\\
	=& \sum_{i=1}^{N}  \sigma_i^{\mu'}\, \left( \MQ_i - \gamma\, \sigma_i^{\mu}\, \MQ_i + \gamma\, \pQ\, (1 - \sigma_i^{\mu})\, \SQ_i \right)\nonumber\\
	&- \pQ \sum_{i=1}^{N} (1-\sigma_i^{\mu'}) \left( \SQ_i - \gamma\, (1-\sigma_i^{\mu})\, \SQ_i + \frac{\gamma}{\pQ}\, \sigma_i^{\mu}\, \MQ_i \right)\\
	=& \sum_{i=1}^{N} \left( \sigma_i^{\mu'}\, \MQ_i - \pQ\,  (1-\sigma_i^{\mu'})\, \SQ_i \right) 
	- \gamma \sum_{i=1}^{N} \left( \sigma_i^{\mu}\, \MQ_i - \pQ\,  (1-\sigma_i^{\mu})\, \SQ_i \right)\\
	=&\ 0 - 0 = 0\\
	&&\Box\nonumber
\end{align}

\subsection{Completeness of the Strategies}

As shown above, a sufficient condition for complete suppression of all price changes is finding a resource distribution $(\MQ, \SQ)$ for which the price is independent of the information. That is, 
\begin{eqnarray}
	p(\mu, \MQ, \SQ) &=& \pQ\ \forall\ \mu
\end{eqnarray}
which is equivalent to \longforref{for:steady_state_condition}. To fulfill this criterion, we need enough agents to form a complete basis in the strategy space which has $D$ dimensions. Then, the deviation from $\pQ$ caused by each agent can be canceled by a superposition of the other agents for every $\mu$. This can be guaranteed if the number of speculators $N_s$ exceeds $2D$.

\begin{figure}
	\includegraphics{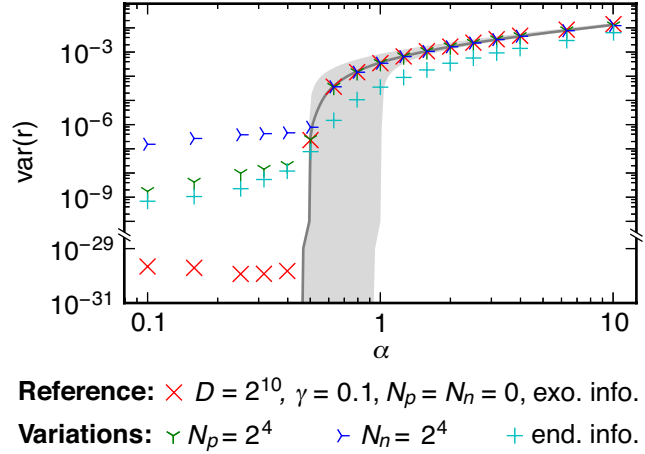}
	\caption{Average log-return variances for different values of $\alpha = D / N_s$. The model with exogenous information and only speculators serves as a reference. For comparison, simulations with a small numbers of either deterministic ($N_p=2^4$) or random producers ($N_n=2^4$) are shown as well as as the model with endogenous information. Grey area: Analytical upper and lower limit for exogenous information. Dark gray line: heuristic interpolation.}
	\label{fig:phase_transition}
\end{figure}

For an insufficient number of speculators, we can still calculate an upper and a lower bound for the variance of the log returns given $D$ and $N$ for a perfect superposition of speculators with exogenous information. Numerical and analytical results for this case are shown in \figref{fig:phase_transition}. The mean variance is found to drop dramatically at $\alpha = D / N_s = 1/2$, with an increasingly sharp transition for large D. This phase transition can be understood by considering the probability that a random binary vector can be canceled by an optimal superposition of $N-1$ random binary vectors with positive weights. As an interim step, consider superpositions of random vectors with arbitrary weights. One such vector creates a one dimensional subspace. Adding a second vector expands the dimensionality of the subspace to $d_2 = 2$ if it is linearly independent of the first one. Adding further vectors one by one, the probability that the $i$th vector does not lie in a $d_{i-1}$-dimensional submanifold is
\begin{equation}
	P(d_i = d_{i-1}+1) = 1 - 2^{D - d_{i-1}}.
\end{equation}
We can therefore iteratively calculate the probability distribution $P(d_{N_s-1})$ of $d$ after adding $N_s - 1$ agents and the probability
\begin{equation}
	P(d_{N_s} = d_{N_s-1}+1) = \sum_{d = 1}^{D} P(d=d_{N_s-1})\ (1 - 2^{d - D})
	\label{for:p_linearindependent}
\end{equation}
that one out of $N_s$ agents is linearly independent of the others. If a vector is linearly independent of the other agents in $d$ dimensions, it cannot be canceled by a linear combination of the other agents for all $\mu$. However, it may still be possible to cancel this agent's impact for a subset of all possible $\mu$, i.e. for a smaller subspace. Therefore, the probability that an agent cannot be canceled in any given time step is
\begin{equation}
	P_\textrm{c.c.} = \sum_{d = 1}^{D} P(d=d_{N_s-1})\ (1 - 2^{d - D})\ (1 - \frac{d}{D}).
	\label{for:p_cant_cancel}
\end{equation}
The last term weights each summand with the fraction of dimensions in which the agent's impact is not canceled. Finally, to relate the fraction of not canceled agents to returns we need to consider the fluctuations prior to any resource redistribution. Since all strategies and $\mu$ are chosen randomly, agents initially contribute to the demand or the supply at random. These fluctuations of demand and supply then follow a binomial distribution with $N_s$ trials and equal probability for buying or selling:
\begin{eqnarray}
	\delta \propto {\cal B}(N_s, 1/2)\\
	\varsigma \propto {\cal B}(N_s, 1/2).
\end{eqnarray}
Since
\begin{eqnarray}
	\langle \delta \rangle = \langle \varsigma \rangle = N_s / 2,\quad \textrm{and}\\
	\var(\delta) = \var(\varsigma) = N_s / 4
\end{eqnarray}
we can approximate the price for small deviations:
\begin{eqnarray}
	p = \frac{N/2 + \Delta\delta}{N/2 - \Delta\varsigma} \approx 1 + 2\frac{\Delta\delta - \Delta\varsigma}{N_s}.
\end{eqnarray}
Therefore,
\begin{eqnarray}
	\langle p(0) \rangle &=& 1,\\
	\var(p(0)) &\approx& 4 / N_s,
\end{eqnarray}
and finally
\begin{eqnarray}
	\langle r(0) \rangle &=& 0,\\
	\var(r(0)) &\approx& \frac{8}{N_s \ln(10)^2},
	\label{for:var(r(0))}
\end{eqnarray}
where the logarithm stems from using the logarithm with base $10$ in the log return.

Combining \eqns \forref{for:p_cant_cancel} and \forref{for:var(r(0))}, we obtain the expected variance of the return for an optimal superposition of agents without the positivity constraint on the resources
\begin{equation}
	\var(r_\textrm{n.c.}) = \var(r(0))\ P_\textrm{c.c.}.
	\label{for:var(r_nc)}
\end{equation}
Since resources cannot be negative, they form a positive cone. Each agent thatlocalized is linear independent of the others spans a half space. Therefore, $2 N_s$ agents are necessary to completely span the strategy space. Yet for small numbers of agents, each agent still represents a full degree of freedom since the probability that two agents lie on the same 1-dimensional submanifold is vanishingly small. However, as the number of agents is increased such that $\alpha \rightarrow 1$, an increasingly large number of new agents just converts a halfspace into a full one. Therefore, \longforref{for:var(r_nc)} represents a lower limit for the variance of the log returns which is a good description for $N_s \ll D$. An upper limit is obtained by changing \longforref{for:p_cant_cancel} such, that each agents increases $d$ by $1/2$. This is a good approximation for $N_s \approx 2 D$. The area in between these limits is shown in \figref{fig:phase_transition} (gray shaded). The lower limit has a phase transition at $\alpha=1$ while the upper limit has a phase transition at $\alpha=1/2$. A phase transition at $\alpha = 1$ is already present in \longforref{for:p_linearindependent}. The gradual convergence for the true variance of the system from the lower to the upper limit is captured by a simple heuristic interpolation: For the dark gray line in \figref{fig:phase_transition}, the probability for a new linearly independent agent to increase $d$ by one is $P_1=\textrm{min}(1, N_s / 2^{m+1}$) while the probability to increase $d$ by $1/2$ is $P_{1/2}=1 - p_1$.
The presented theory describes the numerical results (\figref{fig:phase_transition}) for the model with endogenous information very well for $\alpha \leq 1/2$. For full markets, the residual error for simulations with only speculators is determined by the numerical precision. When producers are present, the residual error is noticeably higher. This is due to the fact, that producers push the system off the invariant manifold. This error is dependent on the agents' use and vanishes for small $\gamma$. Still, predictable producers are canceled much better than random ones because speculators can successfully predict their choices. For endogenous information ($D = 2^K$), the phase transition appears smother and slightly shifted towards larger $\alpha$. A stronger reduction of average returns for $\alpha < 1/2$ occurs due to the more localized adaptation: Agents do not adapt to all possible values of $\mu$ at the same time. 

\subsection{Gradient Descent}
\label{sec:gradient}

We now investigate, how the system evolves towards the invariant manifold. We focus on large numbers of agents and small $\gamma$. The resource redistribution due to subsequently trading the two assets for one another is found to be a special case of a learning rule which minimizes log-return magnitudes. Even more generally, we consider the error function
\begin{equation}
	e = r^2
\end{equation}
and show that its gradient
\begin{equation}
	\frac{\partial e}{\partial X} = 2 r \frac{\partial r}{\partial X}, \quad X \in \{M, S\}
	\label{for:gradient}
\end{equation}
with respect to the agents' resources is dominated by terms with the opposite sign as the change in the agents' resources. Therefore, any scaling of the agents' resources which keeps the sign of the return for money and the opposite sign for stocks corresponds to minimizing log return magnitudes similar to a gradient descent. 

To begin with, consider two subsequent time steps where the information takes the values $\mu$ and $\mu'$. We again consider a market  consisting of speculators only. The derivative of the return with respect to the resources of an agent $k$ is
\begin{eqnarray}	
		\label{for:drdm}
\frac{\partial r(M, S, \mu, \mu')}{\partial M_k} &=& \frac{\sigma_k^{\mu'}}{\delta'} - \frac{\sigma_k^{\mu}}{\delta} + O(\gamma),\\
	\frac{\partial r(M, S, \mu, \mu')}{\partial S_k} &=&  \frac{1 - \sigma_k^{\mu}}{\varsigma} - \frac{1-\sigma_k^{\mu'}}{\varsigma'} + O(\gamma),\quad \textrm{with}\\[\baselineskip]
	\delta = \delta(M, \mu), && \delta' = \delta(M', \mu'),\\
	\varsigma = \varsigma(S, \mu), && \varsigma' = \varsigma(S', \mu').
\end{eqnarray}
The change in resources after trading twice is
\begin{eqnarray}
	\Delta M_k = M_k'' - M_k &=& \gamma\left( S_k ((1-\sigma_k^\mu) p + (1-\sigma_k^{\mu'}) p') - M_k (\sigma_k^{\mu'} + \sigma_k^\mu)\right) + O(\gamma^2)\\
	\Delta S_k = S_k'' - S_k &=& \gamma\left( M_k \left(\frac{\sigma_k^{\mu'}}{p'} + \frac{\sigma_k^\mu}{p}\right) -  S_k (2 - \sigma_k^\mu - \sigma_k^{\mu'} )\right) + O(\gamma^2).
\end{eqnarray}
We are interested in
\begin{equation}
 	\Delta r_k = \left(\Delta M_k \frac{\partial r}{\partial M_k} + \Delta S_k \frac{\partial r}{\partial S_k} \right)
\end{equation}
and continue only with leading terms in $\gamma$. 

For now, we also assume that agents can only perform roundtrip trades (RT). The general case will be discussed later. Two cases are left:
\newline
Case $\sigma_k^{\mu} (1 - \sigma_k^{\mu'}) = 1$:
\begin{eqnarray}
	\frac{\Delta r_k^{RT}}{\gamma} &\stackrel{\gamma \ll 1}{\approx}& \frac{M_k - p' S_k}{\delta} - \frac{M_k / p - S_k}{\varsigma'}\\
	&=&  \frac{M_k}{\delta} (1-\frac{\varsigma}{\varsigma'}) + \frac{S_k}{\varsigma'} (1-\frac{\delta'}{\delta})
\end{eqnarray}
Case $\sigma_k^{\mu'} (1 - \sigma_k^{\mu}) = 1$:
\begin{eqnarray}
	\frac{\Delta r_k^{RT}}{\gamma} &\stackrel{\gamma \ll 1}{\approx}&  \frac{p S_k - M_k}{\delta'} + \frac{M_k / p' - S_k}{\varsigma'}\\
	&=&  \frac{M_k}{\delta'} (\frac{\varsigma'}{\varsigma}-1) + \frac{S_k}{\varsigma} (\frac{\delta}{\delta'}-1).
\end{eqnarray}
Above, we used $p = \delta/\varsigma$ and $p' = \delta'/\varsigma'$. Then,
\begin{eqnarray}
	\frac{1}{\gamma} \sum_{k=1}^{N} \Delta r_k^{RT} &=& \frac{\varsigma'}{\varsigma} - \frac{\varsigma}{\varsigma'} + \frac{\delta}{\delta'}- \frac{\delta'}{\delta}\\
	&=& \frac{\varsigma'}{\varsigma} (1 - \frac{p'}{p}) + \frac{\varsigma}{\varsigma'} (\frac{p}{p'} - 1)\quad
	\left\{ \begin{array}{lcl}
		< 0, \quad r > 0\\
		> 0, \quad r < 0\quad .
	\end{array}\right.
	\label{for:sum_delta_r}
\end{eqnarray}
Therefore, the change in the total error function
\begin{eqnarray}
 	\sum_{k=1}^{N}  \left(\Delta M_k \frac{\partial r^2}{\partial M_k} + \Delta S_k \frac{\partial r^2}{\partial S_k} \right) \stackrel{\textrm{RT}}{\leq} 0
\end{eqnarray}
can never be positive if agents only perform roundtrip trades. 

On average, this result holds even for the general case. The reason why we have to consider averages is, that agents who buy or sell two times in a row always decrease the amount of money or stocks they own after two time steps. Therefore, these agents' resources are expected to change in the opposite direction of the gradient half of the time. That is, for every given pair of informations $(\mu, \mu')$, a quarter of all agents' is expected to have their resources evolve such that future $r(\mu,\mu')^2$ increase. However, the actual influence of these agents is much lower and can be neglected for large systems. Then, demand and supply can be well described as binomial processes as shown above. Here, express demand and supply as:
\begin{eqnarray}
	\delta =& \frac{N}{2} + \Delta\delta,\quad \langle\Delta\delta\rangle &= 0,\quad \langle\Delta\delta^2\rangle \leq \frac{N}{4}\\
		\varsigma &= \frac{N}{2} + \Delta\varsigma,\quad \langle\Delta\varsigma\rangle &= 0,\quad \langle\Delta\varsigma^2\rangle \leq \frac{N}{4}
\end{eqnarray}
and the relative fluctuations around the mean demand $N/2$ are only $\sqrt{N}/2$ and therefore small for large N. Thus we can expand \longforref{for:drdm} for small fluctuations:
\begin{eqnarray}
	\frac{\partial r(M, S, \mu, \mu')}{\partial M_k} &\approx& \sigma_k^{\mu'} \left( \frac{2}{N} - \frac{4 \Delta\delta'}{N^2} \right) - \sigma_k^{\mu} \left( \frac{2}{N} - \frac{4 \Delta\delta}{N^2} \right)\\
	\frac{\partial r(M, S, \mu, \mu')}{\partial S_k} &\approx& (1-\sigma_k^{\mu}) \left( \frac{2}{N} - \frac{4 \Delta\varsigma}{N^2} \right) - (1-\sigma_k^{\mu'}) \left( \frac{2}{N} - \frac{4 \Delta\varsigma'}{N^2} \right).
\end{eqnarray}
As the above equation shows, when agents perform roundtrip trades, they contribute a term of order $N^{-1}$ to the gradient with respect to each asset. When agents buy or sell twice, they only contribute a term of order $N^{-1.5}$ for one asset. Therefore, the influence of these agents vanishes for sufficiently large $N$. By a similar argument, approximately a quarter of all agents performs either one of the actions (buy, sell), (sell, buy), (buy, buy), and (sell, sell) while fluctuations can be neglected for large $N$. Summing up, the expected change in $r(\mu, \mu')^2$ over repeated trades with the same information
\begin{eqnarray}
 	\left\langle \sum_{k=1}^{N}  \left(\Delta M_k \frac{\partial r^2}{\partial M_k} + \Delta S_k \frac{\partial r^2}{\partial S_k} \right)\right\rangle_{(\mu, \mu')} \leq 0.
\end{eqnarray}
is always negative given a sufficiently large number of agents.

\newpage
\section{Supplementary Figures}
\subsection{Phase Diagrams for Exogenous Information}

\begin{figure}[h]
	\includegraphics[trim=3cm 0 0 0]{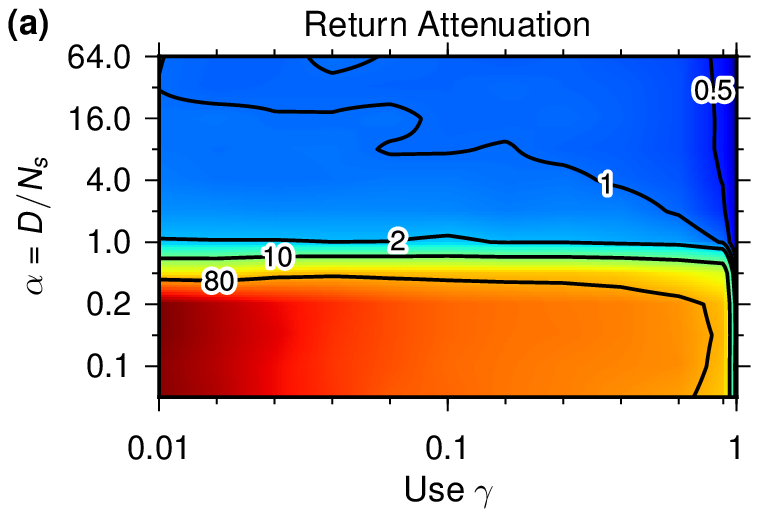}
	\includegraphics[trim=0 0 3.8cm 0]{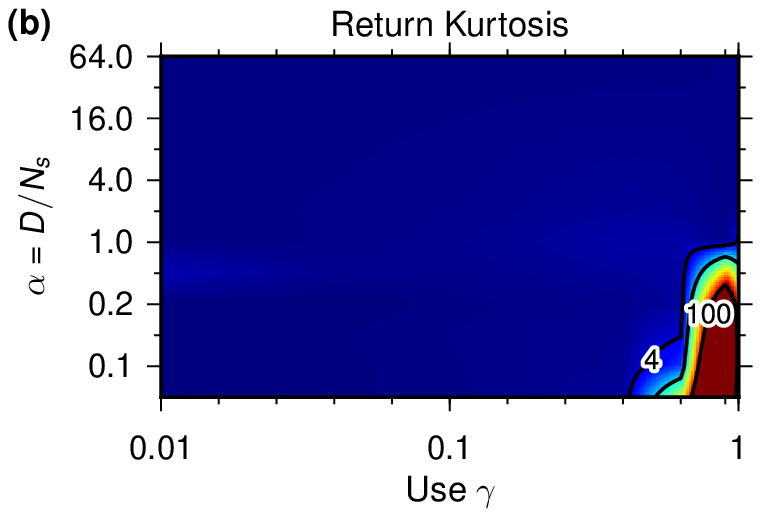}
	\caption{\textbf{(a)}: Reduction of average return magnitudes during transients, and \textbf{(b):} kurtoses for the model with exogenous $\mu$ drawn from a uniform distribution. All other parameters and figure generation are identical to \figref{fig:use_vs_bet} in the main paper.}
	\label{fig:use_vs_bet_ext}
\end{figure}

\noindent For uniformly distributed exogenous information states, a reduction of fluctuations does not generally increase the kurtosis (\figref{fig:use_vs_bet_ext}). This is opposed to the endogenous case shown in \figref{fig:use_vs_bet} in the main paper. Only very large uses in overcomplete markets cause high kurtoses. Then, even uniformly distributed $\mu$ occasionally are not repeated for a sufficiently long time to be ``forgotten'' by the market.\\[2\baselineskip]

\newpage
\subsection{Speculators and Producers}

\begin{figure}[h]
	\includegraphics[trim=3cm 0 0 0]{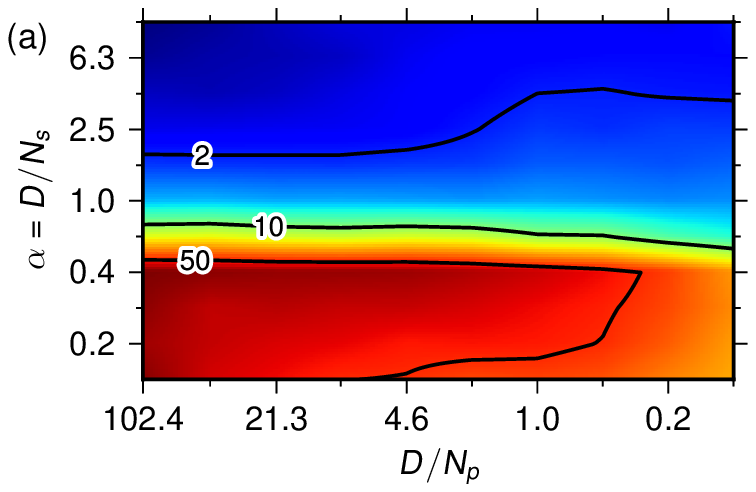}
	\includegraphics[trim=0 0 3.8cm 0]{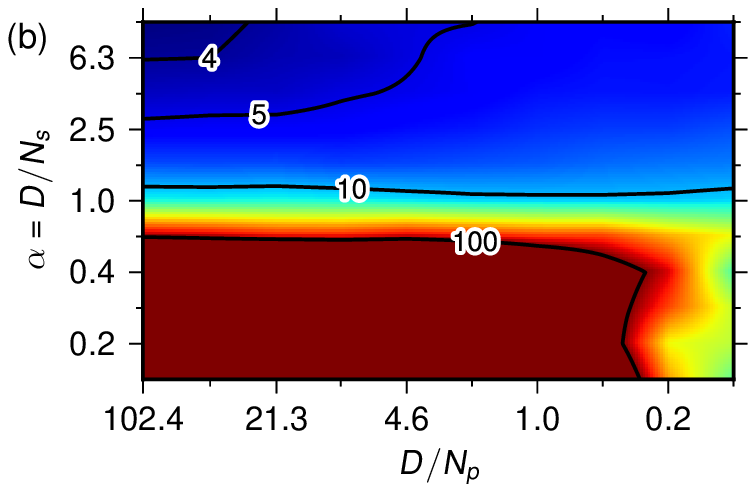}
	\caption{\textbf{(a)}: Reduction of average return magnitudes during transients, and \textbf{(b):} kurtosis of log-returns of the model with endogenous information for different numbers of speculators $N_s$ and producers $N_p$ for constant memory $K=~10$ (i.e. $D=2^{10}$) and use $\gamma = 0.5$. 
	}
	\label{fig:chartis_vs_fundis}
\end{figure}

\noindent \figref{fig:chartis_vs_fundis} shows the phase diagram for $\alpha$ versus the amount of producers in the market. As it turns out, a second phase transition with respect to the number of producers is found. This transition is independent of the one for the speculators. Small $N_p < 0.5 \cdot D$ only weakly influence return distributions. \\[2\baselineskip]

\newpage
\subsection{Distribution of Information Ages (Surprise)}

\begin{figure}[h]
	\includegraphics{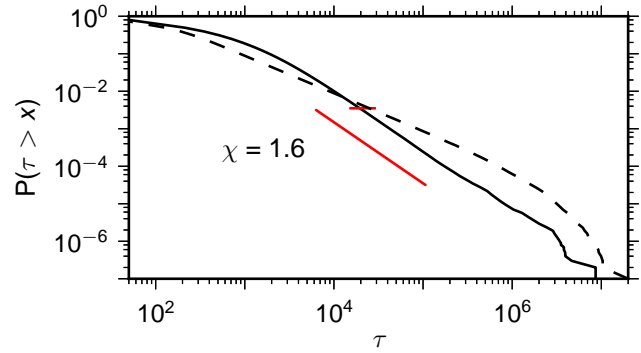}
	\caption{Complementary cumulative distribution of the times $\tau(t)$ since the informations $\mu(t)$ occurred last. Solid black line: Model with intrinsic information and $D = 2^{10}$, $N_s = 2^{11}$, $N_p = 0$, $\gamma = 0.5$. Short red line: Hill estimator. Dashed line: Exogenous information with $P_\textrm{exo}(\mu) \propto \exp(-0.02\, \mu)$, leading to $P(\tau) \propto \tau^{-2}$. }
	\label{fig:surprise_distribution}
\end{figure}

\newpage
\subsection{Mixed Information}

\begin{figure}[h]
	\includegraphics[trim=3.6cm 0 0 0]{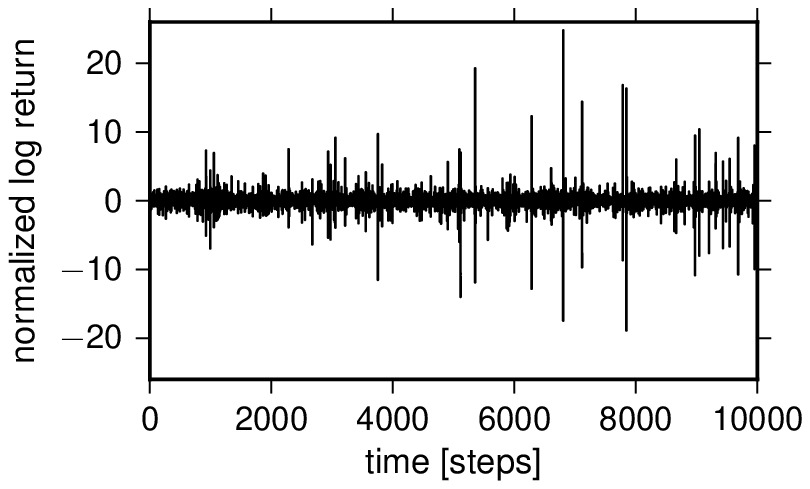}
	\includegraphics[trim=0 0 3.8cm 0]{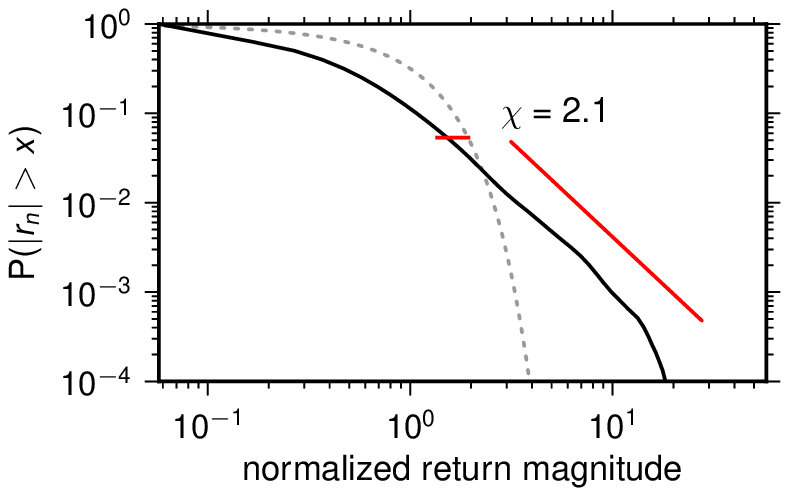}
	\caption{Simulation for a model with $3$ bits of uniform exogenous information and $6$ bits of endogenous information. $N_s = 2^{10}$, $N_p = 0$, $\gamma = 0.1$. The first $2^7$ time steps were discarded. Log returns $r_n$ are normalized by their standard deviation. \textbf{(a)}: Time series \textbf{(b):} Solid black line: complementary cumulative distribution function. Short Red line: Hill estimator for the scaling exponent. Dashed grey line: normal distribution. }
	\label{fig:mixed_info}
\end{figure}

\noindent For a combination of endo- and exogenous information, results are similar to pure endogenous information as long as the endogenous part dominates. Generally, more exogenous information leads to a stronger reduction of fluctuations, less pronounced volatility clustering, and random time series without visible patterns even for small $\gamma$. The scaling of the remaining extreme returns remains unchanged. An example is shown in \figref{fig:mixed_info}. \\[2\baselineskip]

\newpage
\section{Supplementary Discussion: Income and the Critical Point}
\label{apdx:income}

\begin{figure}
	\includegraphics{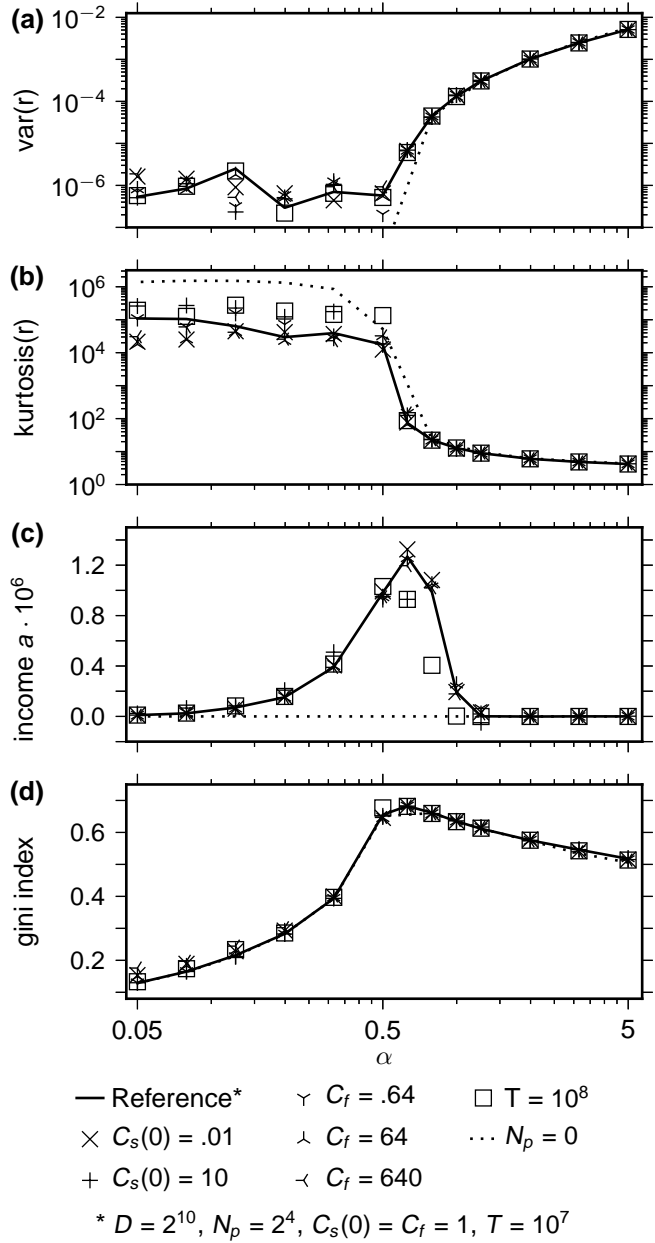}
	\caption{Properties of log-return and resource distributions depend on the parameter $\alpha = D / N_s$. Solid line: The model with unity (initial) speculator and producer capitals $C_s$ and $C_p$ and a small number of producers serves as a reference. Shown are averages from $50$ simulations for each of which only the last $T = 10^7$ out of $2\,T$ timesteps have been analyted. \textbf{(a)}: variances and \textbf{(b)}: kurtoses were calculated from log-returns. \textbf{(c)}: income factors $a$ according to \eqns \forref{for:totalcapital} and \forref{for:income}.  \textbf{(d)} gini indices for speculator capitals after $2\,T$ timesteps.
	For each other line or symbol, only one model parameters has been changed, respectively.
	}
	\label{fig:criticalpoint}
\end{figure}

\noindent \figref{fig:criticalpoint} shows the phase transition with respect to $\alpha$ in more detail. As in \figref{fig:phase_transition}, we take one parameter set as a reference to which we compare simulations after transients for different parameters. For orientation, the log-return variances (\subfigref{fig:criticalpoint}{(a)}) and kurtoses(\subfigref{fig:criticalpoint}{(b)}), which have been discussed earlier, are shown again.

\noindent Mean speculator capitals
\begin{equation}
	C_s(t) = \frac{1}{2 N_s} \sum_{k=1}^{N_s} M_k(t) + S_k(t).
	\label{for:totalcapital}
\end{equation}
are not constant over time in markets that include producers. For empty markets, the ratio of average speculator and producer capitals quickly evolves towards an equilibrium. The more agents are added, the longer it takes for $C_s$ to saturate. For critical or crowded markets a positive speculator income persists over long times. Then, average speculator capitals after transients can be well described as:
\begin{equation}
	C_s(t)^2 = C_s(t_0)^2 + a\, t, \quad t_0 < t.
	\label{for:income}
\end{equation}
The income factor $a$ is shown in \subfigref{fig:criticalpoint}{(c)} and quantifies how well the speculators can exploit the producers. $a$ is found to be independent of the initial ratio between speculator and producer capitals. It becomes maximal close to the critical point which can be intuitively understood: For empty markets, there is a finite chance for a producer strategy to lie outside of the space spanned by the speculators. Therefore, increasing the number of speculators increases their average income. For crowded markets, producers are already optimally exploited. Then, adding more speculators just distributes the maximal total income over more of them. An analogous maximum can be found in Minority Games (see e.g. ref. 15 in the main paper).

\subfigref{fig:criticalpoint}{(d)} shows the gini index, a common measure of wealth inequality. Increased incomes coincide with increased capital inequality among speculators as the gini-index shows a maximum at the critical point. Thus, only few speculators are most successful in exploiting the producers.

\end{document}